\documentclass[onecolumn]{aa}
\usepackage{graphicx}
\begin{document}
   \title{Properties of the linearly polarized radiation
          from PSR B0950+08 }

   \author{T. V. Shabanova \and Yu. P. Shitov}

   \offprints{T. V. Shabanova}

   \institute{Astro Space Center, P. N. Lebedev Physical Institute,
              Leninskij Prospect 53, 117924 Moscow, Russia  \\
              \email{tvsh@prao.psn.ru, shitov@prao.psn.ru}
             }

   \date{Received 30 July 2002 / Accepted 26 November 2003 }

   \abstract {
   Measurements of average pulse profiles made with a single
   linear polarization over the range 41--112 MHz are presented for
   PSR B0950+08. We show that the observed variable structure of
   the pulse profiles is a result of Faraday sinusoidal modulation
   of the pulse intensity with frequency.
   The rotation measure corresponding to this effect,
   $RM \approx 4\ {\mathrm{rad\, m^{-2}}}$, is about 3 times greater
   than the published value of $RM = 1.35\ {\mathrm{rad\, m^{-2}}}$
   (Taylor et al. \cite{taylor}).

              }

   \titlerunning{Linearly polarized radiation from PSR B0950+08}
   \authorrunning{Shabanova \and Shitov}

   \maketitle
%

\section{Introduction}

   PSR B0950+08 is well-studied over a wide
   frequency range from 24 to 10500 MHz. It exhibits a single
   pulse profile at frequencies above 400 MHz and a double
   pulse profile at low frequencies in the range 24--112 MHz
   (Hankins et al. \cite{hankins}; Kuzmin et al. \cite{kuzmin}).
   There is an interpulse occurring approximately $152{\degr}$
   ahead of the main pulse and a bridge of emission between the
   interpulse and main pulse (Lyne \& Rickett \cite{lyne1}).

   The study of average pulse profiles from a large number
   of pulsars has shown that a pulse profile obtained
   by averaging several hundred individual pulses is very
   stable for most pulsars (Helfand et al. \cite{helfand};
   Rathnasree \& Rankin \cite{rathnas}). Profile changes can be
   caused by such phenomena as mode changing and nulling
   which occur in a small fraction of pulsars.
   The narrowband profile changes which are observed for the pulsar
   B0950+08 in the range 41--112 MHz and discussed in this paper
   are consistent with the effects of Faraday rotation
   on a linearly polarized pulse emission received by a linearly
   polarized antenna. Pulsars with considerable linear
   polarization observed with a single linear polarization
   show frequency-dependent profiles because of Faraday rotation
   of the plane of polarization in the interstellar medium.
   Pulse profiles obtained for different epochs show
   time-dependent profile shapes caused by a variation of the
   electron density or magnetic field in the propagation path,
   mainly due to varying ionospheric contribution to
   the rotation measure.
   At the Pushchino Radio Astronomy Observatory (PRAO),
   the effect of Faraday modulation of the pulse intensity with
   frequency at the output of the multi-channel receiver is used
   for the measurements of linear polarization  characteristics
   of pulsars and the estimation of their rotation measure
   (Vitkevich \& Shitov \cite{vitkev}; Shitov \cite{shitov1};
   Suleymanova \cite{suleym}).

   Changes of the average pulse profile at meter wavelengths
   for the pulsar B0950+08 were first noticed
   by Smirnova \& Shabanova (\cite{smirnova}).
   They found that the pulse profile is variable in time
   and shows narrowband changes with frequency.
   The explanation of the observed phenomenon by Faraday rotation
   was problematic because the predicted effect of the interstellar
   Faraday rotation for the tabulated value of
   $RM=1.35\ {\mathrm{rad\, m^{-2}}}$ (Taylor et al. \cite{taylor})
   was negligible across the receiver bandpass at 102 MHz.
   The authors supposed that the narrowband changes of the pulse
   profile may be due either to the pulsar's intrinsic narrowband
   emission, manifesting only at low frequencies, or to scintillations
   of spatially separate sources. However, the low time resolution of
   their observations (about 10$\degr$ of longitude) made a detailed
   analysis of this phenomenon impracticable.

   The main purpose of this paper is to explain
   the observed narrowband profile changes of PSR B0950+08.
   The techniques used for observations over the frequency range
   41--112 MHz  are described in \S2.
   In \S3 a large set of average profiles is investigated
   with respect to the shape changes with time.
   In \S4 we study the narrowband changes of the pulse profile
   across the 2.56-MHz bandpass and establish a relation between
   the shape changes with time and the shape changes with frequency.
   \S5 discusses the frequency dependence of the pulse profile
   after removing the effect caused by interstellar scintillation.
   In \S6 we compute Faraday rotation effects using
   a numerical polarization model of the pulse profile.
   \S7 describes the narrowband changes of the average profile
   at the lower frequencies of 88, 62 and 41 MHz.
   \S8 presents the results of the timing data analysis for
   the pulsar. In \S9 we conclude that
   the profile changes are due to the Faraday rotation effect.


\section{Observations}

   Observations with a high time resolution of average
   pulse profiles and individual pulses from the pulsar
   B0950+08 were carried out at PRAO in three observing sessions
   from April 1996 to May 2002.

   The first session took place from April 1996 through July 2001
   and included regular timing observations of the pulsar with
   the BSA radiotelescope which operated at 102.7 MHz until
   May 1998, and at 111.3 MHz since November 1998. The BSA large
   phased array radiotelescope, making up a linearly polarized
   transit antenna with 30000 ${\mathrm{m^{2}}}$ effective area
   and a beam size of about $(3.5\, / \cos \delta$) minute,
   provides 3.3 minutes duration observations at the pulsar
   declination $\delta \approx 8{\degr}$.
   A 32 channel ${\times}$ 5 kHz spectrometer covering a total
   bandwidth of 160 kHz was used to provide  high frequency
   resolution. The receiver time constant was 0.3 ms. The data were
   sampled at intervals of 0.2048 ms.
   An average pulse profile in each channel was formed by
   synchronous averaging of 770 individual periods with a predicted
   topocentric pulsar period. After dispersion removal
   all the channel profiles were summed to form an averaged pulse
   profile for a single observation. A narrow bandwidth of one 5-kHz
   channel limited the pulse broadening from the interstellar
   dispersion to 0.09 ms. The observation window width was
   200$\degr$ of longitude.  A total of 264 average pulse profiles
   were derived at two observing frequencies
   around 102.5 and 112 MHz for a 5-yr span of observations.

   The second session included observations of individual pulses
   and was made using the BSA telescope at an observing frequency
   of 111.87 MHz during 2001 and 2002. The 128 spectral channels,
   each of bandwidth 20 kHz, were used to record individual
   pulses for 3.3 minutes of the BSA transit time.
   The receiver time constant was 1 ms and the sampling interval
   was 0.512 or 0.8192 ms. The signal dispersion through one 20-kHz
   channel causes pulse broadening of 0.35 ms.

   The third session was carried out using the East-West arm
   of the DKR-1000 radiotelescope at three lower frequencies
   of 88.57, 62.15 and 41.07 MHz in late 2001 and early 2002.
   Individual pulses were recorded during 16 minutes of the DKR
   transit observation time with the 128 channel ${\times}$ 20 kHz
   spectral radiometer. The parameters of the DKR observations
   are listed in Table~\ref{paramdkr}.
   The effects of the interstellar dispersion in the total
   bandwidth of 2.56 MHz at three observing frequencies were
   91, 263 and 911 ms, respectively.
   The dispersion time delay at 62.15 and 41.07 MHz was
   more than the 253-ms pulsar period.
   To compensate for this time delay, the signal was recorded
   blocks of 5 full periods at 62.15 MHz and blocks
   of 12 full periods at 41.07 MHz. From each block, the signal
   was only analyzed for either the 3 or 8 first successive periods
   at each observing frequency, respectively.
   The average profile was formed by addition of 2200
   pulse periods at 62.15 MHz and 2160 pulse periods at 41.07 MHz.

   \begin{table}
      \caption[]{Parameters of the DKR observations}
         \label{paramdkr}
    $$
         \begin{tabular}{llll}
            \hline
            \noalign{\smallskip}
Parameter                         & \multicolumn{3}{l}{Value}    \\
            \noalign{\smallskip}
            \hline
            \noalign{\smallskip}
Observing frequency (MHz)         & 88.57   & 62.15   & 41.07    \\
Receiver time constant (ms)       & 1       & 1       & 3        \\
Sampling interval (ms)            & 0.8192  & 0.8192  & 2.9696   \\
One bandwidth   (MHz)             & 0.02    & 0.02    & 0.02     \\
Dispersion pulse broadening (ms)  & 0.7     & 2       & 7        \\
Total bandwidth   (MHz)           & 2.56    & 2.56    & 2.56     \\
Dispersion time delay (ms)        & 91      & 263     & 911      \\
Number of pulses averaged         & 3770    & 2200    & 2160     \\
            \noalign{\smallskip}
            \hline
         \end{tabular}
   $$
   \end{table}

   The pulsar B0950+08 shows a high degree of linear polarization
   of about 90$\%$ and a smooth change in the position angle
   through the interpulse and main pulse, as measured
   at 150 and 151 MHz
   (Schwarz \& Morris \cite{schwarz}; Lyne et al. \cite{lyne}).
   Both the BSA and DKR radiotelescopes receive only one
   polarization.
   For our observations, the published value of a rotation  measure of
   $RM=1.35\ {\mathrm{rad\, m^{-2}}}$ (Taylor et al. \cite{taylor})
   predicts negligible effect of Faraday rotation across the 2.56-MHz
   bandpass.
   The expected periods of Faraday modulation of the pulsar emission
   with frequency are 18 MHz at the observing frequency
   of 112 MHz, 14 MHz at 102.7 MHz, 9 MHz at 88.6 MHz,
   3 MHz at 62 MHz and 0.9 MHz at 41 MHz.
   This effect produces a modulation of the pulse emission
   spectrum of less than 15$\%$ at the frequencies within
   the 102--112 MHz range and is significant only at two lower
   frequencies of 62 and 41 MHz.
   The influence of a varying ionosphere on our measurements
   for different days is also small. Most observations of
   the pulsar were made at night time or close to it
   (during autumn or winter), when an ionospheric
   contribution to the rotation measure $RM$ is estimated
   to be less than 0.4 ${\mathrm{rad\, m^{-2}}}$ for the PRAO site
   (Udaltsov \& Zlobin \cite{udaltsov}).

\section{The changes of the average profile with time
         in the range 102--112 MHz}

\begin{figure}
   \centering
   \includegraphics[width=9cm]{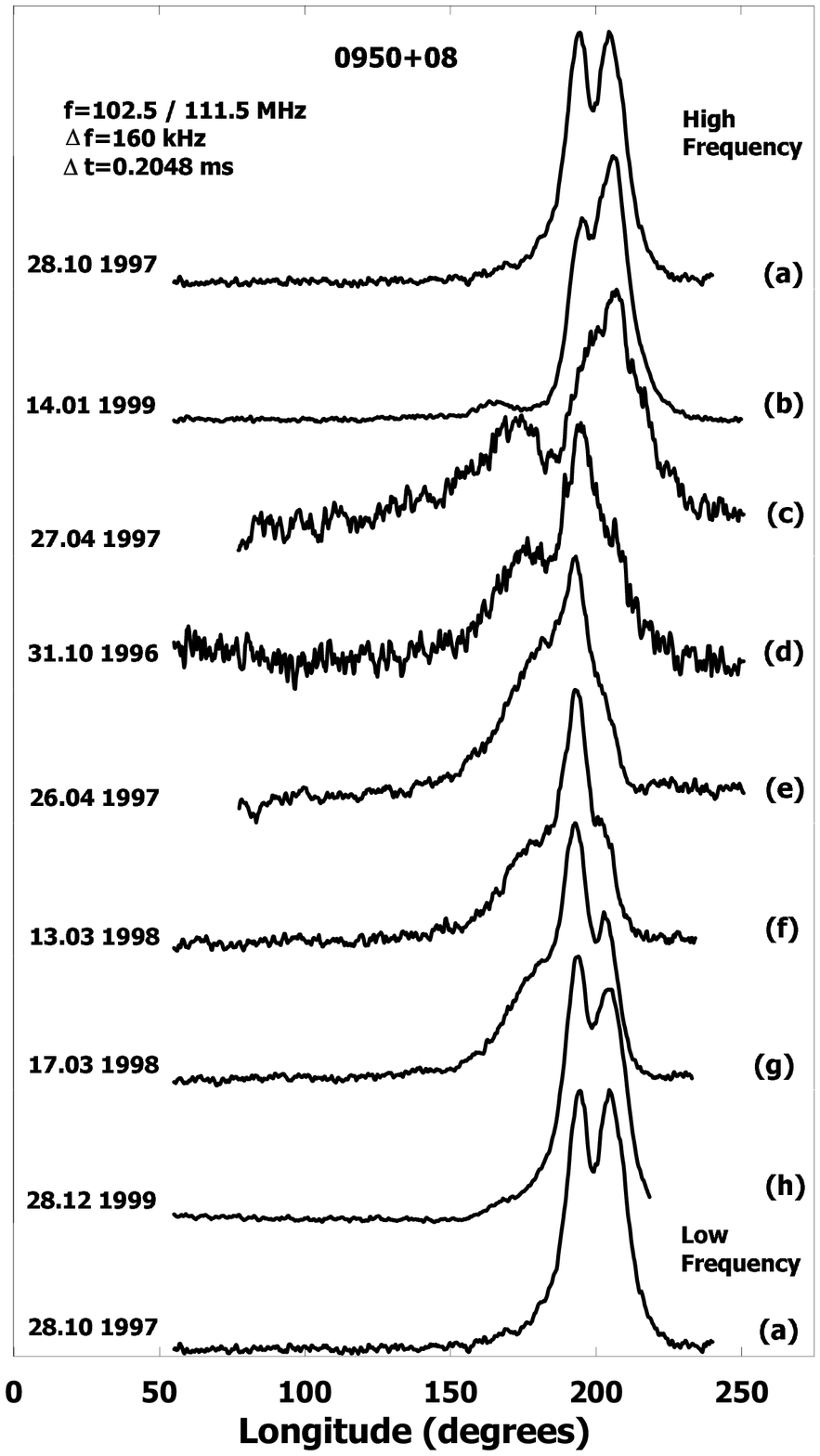}
   \caption{
    A sequence of shapes of the average profile continuously
    varying between the main states of the double profile.
    The profiles were chosen from the full data set obtained
    between 1996 and 2001 at frequencies around 102 and 112 MHz.
    Shapes from $\bf{(b)\ to\ (g)}$ show a triple profile
    with a well-resolved component ahead of the main pulse
    located at longitudes from 150$\degr$ to 180$\degr$.
    The sample interval is $0{\fdg}3$ of longitude.
    All the profiles are aligned in time and normalized to unity.
          }
   \label{prof5khz}
   \end{figure}

   We investigated average profiles using timing observations
   over the interval 1996--2001.
   The shape of the observed pulse profiles
   strongly varies from one observation to another that
   we can see both double and triple profiles.
   To reveal some regularities in the occurrence of
   the profiles with different shapes, we sorted
   available average profiles into groups, each
   containing similar profile shapes.
   Each of the 264 observed profiles could be put into one
   of the profile groups. A detailed analysis
   of these data showed that the shapes from different groups
   make up a set smoothly varying between the main states of
   the double profile. Such a set of continuously varying
   shapes is presented in Fig.~\ref{prof5khz}.

   The shape (a) represents the main pulse, which
   has two well-resolved components with a separation of 10$\degr$.
   The shapes from (b) to (g) show the well-resolved third component,
   occurring $34{\degr}$ ahead of the main
   pulse and visible at a longitude of about 170${\degr}$.
   Note that the unresolved component ahead of the main pulse
   in the range 151--430 MHz was observed earlier
   (Lyne et al. \cite{lyne}; Hankins \& Cordes \cite{hankins1}).
   The shape changes affect the whole
   profile and both the components of the main pulse alter their
   amplitude, either amplifying or weakening alternately.
   Variations in the relative amplitudes of all three components
   exhibit a periodic character.

   The two profiles (c) and (d) of
   Fig.~\ref{prof5khz} have a lower signal-to-noise ratio (S/N)
   and show unusual shapes. These profiles are similar
   to the adjacent profiles (b) and (f) having a better S/N ratio.
   Shapes (c) and (d) are of great importance to confirm
   the periodic character of changes in the pulse profile.
   These shapes provide a transition between the first and
   second dominant components of the main pulse. The state (c)
   where the peak of the second component is significantly higher
   than the peak of the first one quickly changes to another
   state (d) where the peak of the first component dominates.

\section{Changes of pulse shapes with frequency}

   Individual pulses from the pulsar were observed at 111.87 MHz
   with the 128 channel ${\times}$ 20 kHz spectral radiometer.
   The high quality data set was analyzed to investigate
   the frequency structure of the average profiles within bands
   of different widths of 640, 160 and 20 kHz.

\begin{figure}
   \centering
   \includegraphics[width=6cm]{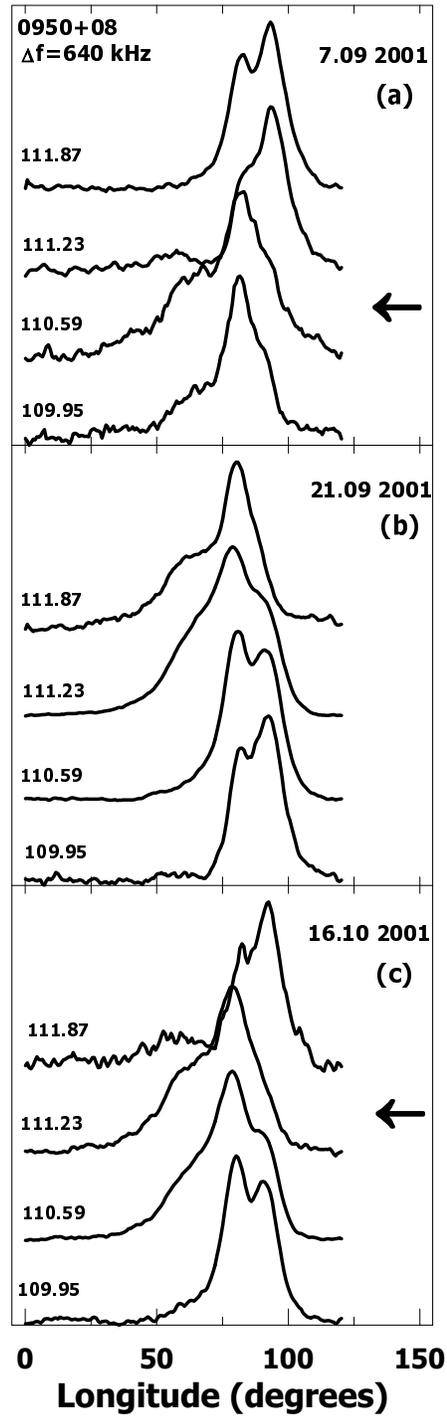}
   \caption{
    The changes of the average profile with frequency for
    three different epochs. A set of 12 plotted
    profiles covers the full cycle of the shape changes.
    Panels $\bf{(a)\ and\ (c)}$ demonstrate the transition
    state between the first and second dominant components of
    the main pulse (marked by arrow).
    The sample interval is $0{\fdg}7$ of longitude.
    All the profiles are normalized to unity.
     }
   \label{prof640}
   \end{figure}

   The pulse profiles were obtained by averaging 770 individual
   periods for each 32-channel quadrant of the filter bank.
   The sequences of 4 resulting profiles are presented in
   Fig.~\ref{prof640} for three different observations.
   Detailed inspection of each panel shows that the shape of
   the average profile changes markedly over the bandpass
   of 2.56 MHz.
   It is also seen that the sequences shown are not identical
   although the same shapes occur in all the cases.
   An analysis of the available data showed that the profile
   shapes vary with frequency in a strictly defined order and
   that the frequency changes are a result of some modulation
   process having a phase drift with time.
   If the process observed is related to Faraday rotation then
   the tabulated value of $RM$ would be expected to be wrong.
   Therefore the predicted Faraday effect cannot be negligible.
   Variable ionospheric Faraday rotation does not affect
   the profiles during the 3 minutes of observation time but causes
   a drift of the frequency picture with time.

   The receiver bandwidth is less than the frequency interval
   of the modulation process and so we observe only some
   stages of this process on different days. Nevertheless,
   a period of the profile frequency changes can be estimated if
   we analyze the records together, which contain
   overlapping shapes as shown in Fig.~\ref{prof640}.
   Here the last shape at 109.95 MHz of the panel~(a) is similar
   to the first shape at 111.87 MHz of panel~(b).
   This is also the case for panels~(b) and~(c). A set of 12 plotted
   profiles covers the full cycle of the shape changes and shows
   how the frequency picture develops from higher
   to lower frequencies, regardless of the particular value of
   the observing frequency.
   The transition between the first and second dominant
   components, which was discussed in the preceding section,
   is clearly observed in panel~(a) between the shapes
   at 111.23 and 110.59 MHz and panel~(c) between
   the shapes at 111.87 and 111.23 MHz. This state indicates
   a transition to a new cycle of profile changes with
   frequency.
   The interval between these states is
   6--7 640-kHz bands, i.e. it is approximately equal to 4 MHz.
   Hence, having a 2.56-MHz band, we can nevertheless estimate
   the frequency interval of the shape changes which can be several
   times wider than a bandwidth.

\begin{figure}
   \centering
   \includegraphics[width=12cm]{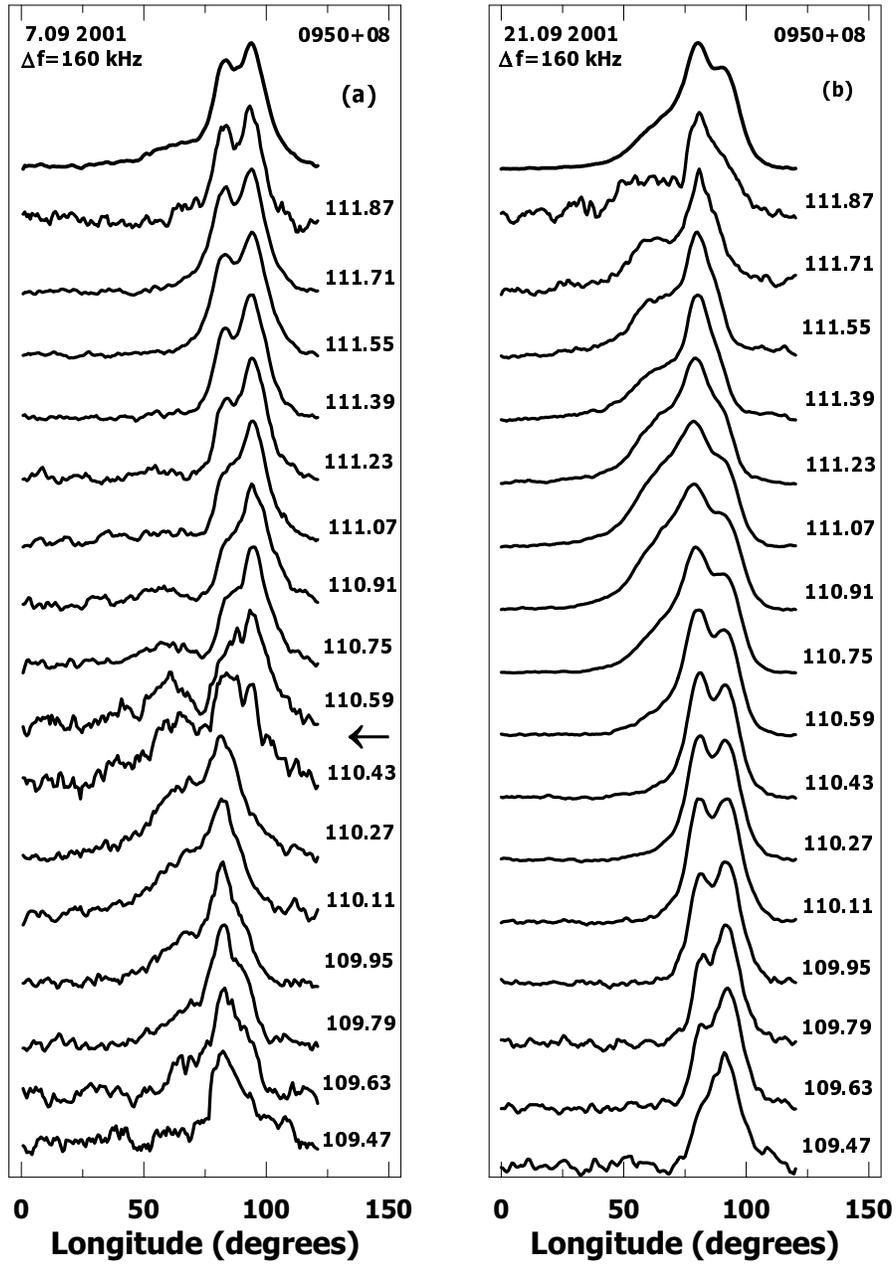}
   \caption{
    The frequency changes of the average profile between
    160-kHz bands for two different epochs.
    To watch the profile frequency dependence inside
    one cycle of changes, the right hand sequence
    should be considered as an extension of the left hand one.
    The transition state between the first and second
    dominant components is marked by the arrow.
    All the profiles are normalized to unity.
         }
   \label{prof160}
   \end{figure}

   For a more detailed analysis of the profile frequency
   structure, individual pulses were averaged within narrower
   160-kHz bands. Two sets of 16 resulting profiles are presented
   in Fig.~\ref{prof160} for two different epochs. They give a finer
   frequency structure of the 640-kHz profiles, already
   displayed in Fig.~\ref{prof640}a~and~b.
   In order to see the profile frequency dependence inside
   one cycle of changes, the right hand sequence
   should be considered as an extension of the left hand one.
   The first component of the profile at 111.87 MHz (panel~(a))
   weakens gradually toward lower frequencies and
   becomes unresolved at 110.59 MHz. Beginning from the profile
   at 110.43 MHz, the first component dominates.
   The last 4 profiles from 109.95 to 109.47 MHz of this
   sequence are similar to the first 4 profiles of the right hand
   sequence. The second component becomes visible in the profile at
   111.23 MHz (panel~(b)). At 110.27 MHz, the two components become
   comparable in amplitude. Then, the first component becomes again
   weaker and the picture is repeated. The profile at 109.79 MHz of the
   panel~(b) is similar to the first profile at 111.87 MHz
   of the panel~(a). It should be noted that the left and right
   sequences show different rates of shape changes with
   frequency. It may be caused by a varying ionospheric contribution
   to the rotation measure on different days.

   We see that the set of the profile shapes plotted in
   Fig.~\ref{prof5khz} is similar to the set of the shapes presented
   in Fig.~\ref{prof160}.
   This suggests that the profile changes with time
   are a result of the profile changes with frequency. When
   the receiver bandpass is narrower than the frequency interval
   of the shape changes, the profile shape obtained will depend on
   the location of the band inside this interval. Note that
   the pulse profiles of Fig.~\ref{prof5khz} were plotted in such
   a way that the sequence of the profile shapes placed from top
   to bottom would correspond to the sequence of the profile
   shapes observed within the receiver bandpass in the direction
   from a higher frequency to a lower one.

   An analysis showed that the shape changes are clearly seen even
   in individual channels of 20-kHz width and these changes
   cannot be caused by instrumental effects.
   The transition state may occupy a bandwidth from 40 to 160 kHz
   for different observations. Weakening of the signal is a typical
   feature for all the observations where this state occurs.

\section{Interstellar scintillations and the frequency
         structure of the profile changes at 112 MHz}

   It is well known that the pulse emission of PSR B0950+08
   is modulated by interstellar scintillations. In order to
   determine the type of frequency dependence of the profile
   changes we must exclude modulation due to scintillations.
   The emission spectrum of the pulsar within the receiver
   bandpass was obtained using the following procedure.
   For each 20-kHz channel out of 128, the area under the average
   profile was calculated on the longitude interval above
   the 0.1 level of the profile peak and normalized to the rms of
   noise outside the profile in the same channel.
   The spectrum of scintillations is characterized by
   modulation of the area under the average profile with
   frequency and is estimated by the width of an autocorrelation
   function $ACF$. For our observations at 112 MHz,
   modulation of the emission spectrum of the pulsar
   is close to 80$\%$. The width of the decorrelation band of
   the spectrum of scintillations defined by the half-width at
   the $(1/e)$ level of $ACF$ is equal to 250 kHz. This agrees well
   with earlier measurements at 105 MHz made by Shitov
   (\cite{shitov2}).

\begin{figure}
   \centering
   \includegraphics[width=12cm]{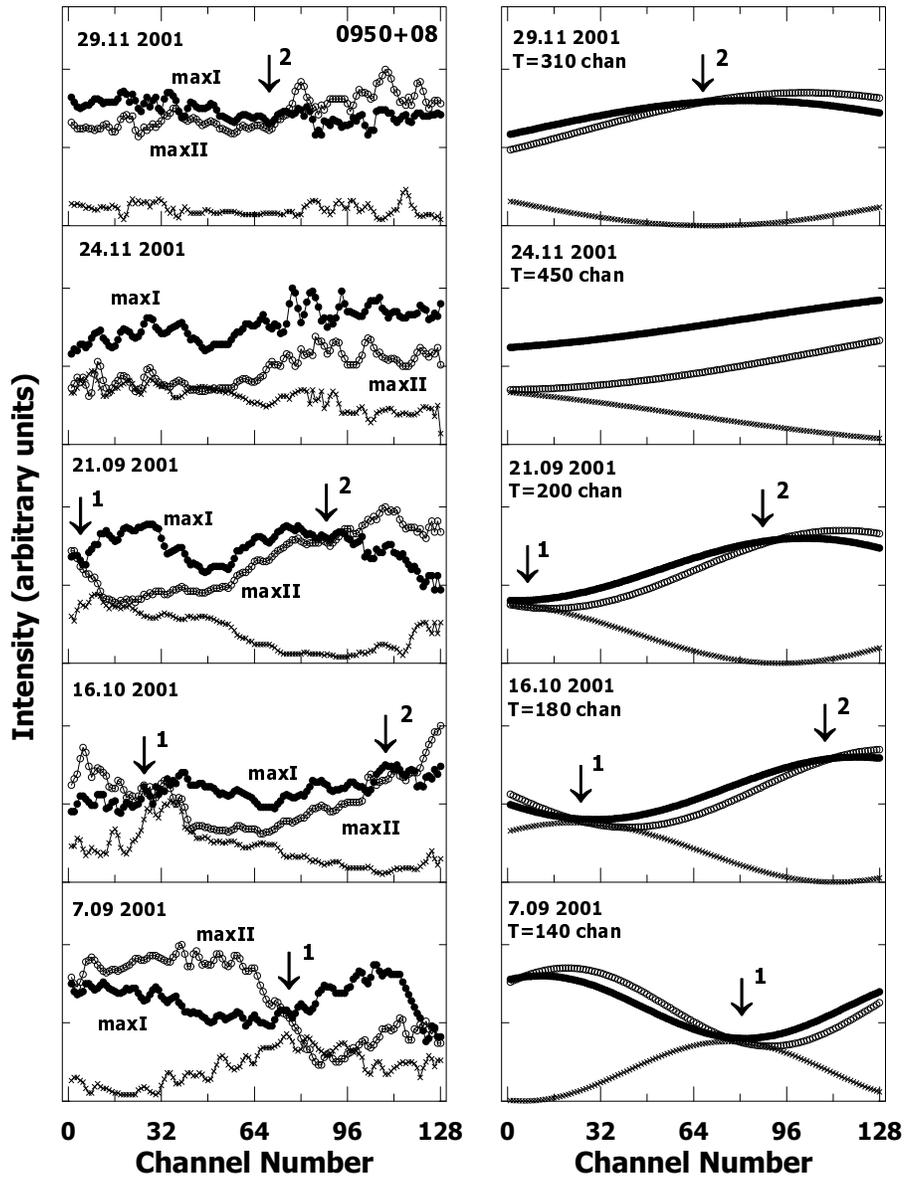}
   \caption{
    The frequency structure of the average profile at
    the longitudes of the pulse peaks of the three components
    for different epochs:
    the observed spectra {\bf{(Left)}},
    the model spectra {\bf{(Right)}}.
    The spectrum for the first component of the main pulse
    is marked by filled circles and the word "maxI",
    the spectrum for the second component is marked by
    open circles and the word "maxII", the spectrum for the third
    component is marked by crosses.
         }
   \label{specpf}
   \end{figure}

   The emission spectrum  of the pulsar in the receiver
   bandpass was calculated by dividing the amplitude of
   the average profile for each channel
   by the area under the average profile in this channel.
   Some examples of the observed emission spectra
   at the longitudes of pulse peaks of the three components
   are given in the left panel of Fig.~\ref{specpf}
   for different observations.

   As is seen in this plot, the spectra of the
   two main components of the pulse profile intersect at two
   points, which correspond to two basic shapes of the average
   profile. Point~1 corresponds to a transition state between
   the first and the second dominant components of the main pulse
   (see triple profiles (c) and (d) in Fig.~\ref{prof5khz}).
   Here the pulse peak of the first component becomes higher
   than the pulse peak of the second component. Near this point,
   the pulse peak of the third component ahead of the main pulse
    has the greatest value.
   Point~2 corresponds to a state when the pulse peaks
   of both the dominant components have comparable amplitudes, and
   the amplitude of the third component approaches its smallest
   value (see double profile (a) in Fig.~\ref{prof5khz}).

   Uniform behavior of the spectral curves for all
   the observations can specify that the narrowband changes
   in the pulse shape are deterministic. A detailed analysis
   of these data showed that the frequency structure
   of the pulse profile can be modelled by a sinusoid.
   We approximated the obtained spectra of the pulse emission
   by three sinusoidal curves, defined for each component as

   \begin{equation}
       y1(k) = 0.20\, \sin\, (\frac{2 \pi}{T}*k + {\phi}) + 0.60
        \label{y1} \,,
   \end{equation}

   \begin{equation}
       y2(k) = 0.25\, \sin\, (\frac{2 \pi}{T}*k - 0.4 + {\phi}) + 0.60
       \label{y2}  \,,
   \end{equation}

   \begin{equation}
       y3(k) = 0.19\, \sin\, (\frac{2 \pi}{T}*k + \pi + {\phi}) + 0.19
       \label{y3}  \,,
   \end{equation}

   where $T$ is the period of the sinusoidal curve measured in
   the channels, ${\phi}$ is a variable phase of the sinusoidal
   curve and $k$ is the channel number.
   The amplitudes of the three approximation curves and
   their constant phase shifts were defined by a method
   of successive approximations. A necessary condition
   was that the cross points of the three sinusoids should correspond
   to the cross points of the observed curves.
   It was assumed that the approximation curve $y(k)$ gives
   a good fit when the goodness-of-fit parameter,
   $S=\sum_{k=1}^{128}[f(k)-y(k)]^{2}$, is minimum.
   Here $f(k)$ is the observed curve. A formal least-squares
   fitting procedure was not applied because it gave unreasonable
   estimations of the parameters in view of the large scattering
   of the experimental data and short segments of the observed curves.

   We applied the approximation procedure to all the cases
   presented in the left panel of Fig.~\ref{specpf}, varying
   the frequency scale $T$ and location $\phi$ of the 2.56-MHz
   band. Initial amplitudes and phases of the curves remained
   the same. The results are presented in the right
   panel of Fig.~\ref{specpf} in the same scale.
   Comparison of the spectra plotted in both panels indicates
   that the suggested model approximates the observed cases
   very well. The profile changes
   at the longitudes of the three components show appreciable
   sinusoidal dependence on frequency. The frequency scale of
   the shape changes varies from 140 to 450 channels
   (or from 3 to 9 MHz) for different observations at 112 MHz.

   A more detailed picture of the amplitude changes
   of the average profile with frequency is shown in
   Fig.~\ref{sin256}. The approximation curves were
   calculated for $T$=256 and  $\phi$=0.
   A sinusoidal modulation covers all the three
   components of the average profile, but at the longitudes
   of each component it happens  with  different phase lags.
   The lags do not depend on time. This produces the profile
   shapes varying with frequency in a similar way for different
   observations.  Fig.~\ref{sin256} also shows that all
   the observed profile shapes can be explained by a different
   location of the receiver bandpass within a cycle of
   the sinusoidal curve.

\section{Numerical polarization model of the average profile}

\begin{figure}
   \centering
   \includegraphics[width=8cm]{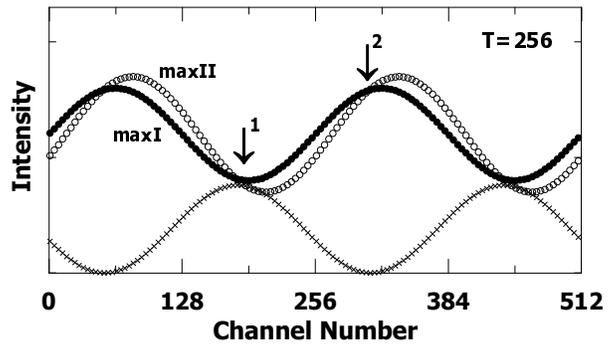}
   \caption{
    The model picture of the amplitude changes of the average
    profile with frequency at the longitudes of the pulse
    peaks of the three components for $T=256$ and $\phi$=0.
    The spectrum is marked by filled circles and the word "maxI"
    for the first component, open circles and the word "maxII"
    for the second component and crosses for the third one.
         }
   \label{sin256}
   \end{figure}

\begin{figure}
   \centering
   \includegraphics[width=13cm]{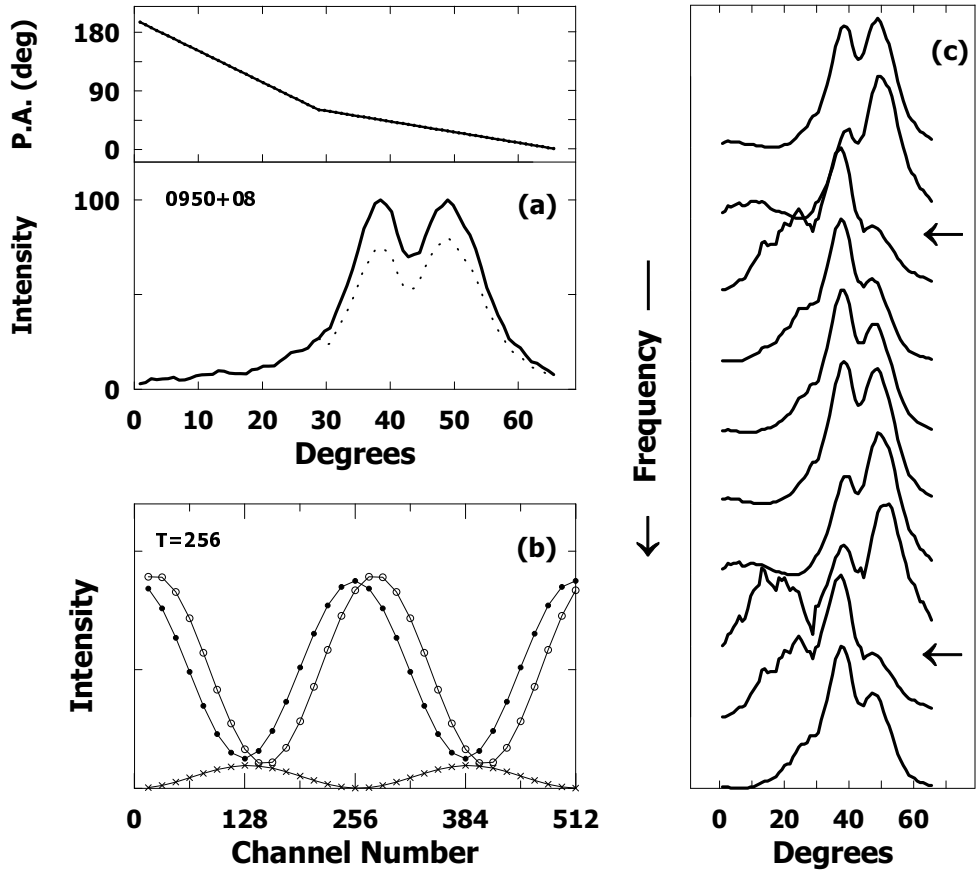}
   \caption{
   A numerical study of polarization effects.
   $\bf{a)}$ The model profile adopted shows average profile
   in total intensity (solid), the linearly polarized
   component (dotted), and the position angle of polarization
   (upper panel).
   $\bf{b)}$ The computed Faraday rotation spectra at the longitudes
   of the pulse peaks of the three components: 14$\degr$ (crosses),
   38$\degr$ (filled circles) and 49$\degr$ (open circles).
   $\bf{c)}$ The computed profile shapes for individual channels.
   All the profiles are normalized to unity.
   A transition state between the first and second dominant
   components of the main pulse is marked by an arrow.
         }
   \label{model}
   \end{figure}

   The frequency picture shown in Fig.~\ref{sin256} is very
   similar to a Faraday rotation pattern. To clarify
   whether the observed effect is due to Faraday rotation,
   we determined a numerical polarization model of the profile
   and computed the effects due to Faraday rotation using
   the technique that is applied for measurements of
   linear polarization characteristics of pulsars at PRAO.
   The frequency-time dependence of a Faraday rotation pattern
   was computed using the equation

   \begin{equation}
       A_{k}(i) = a(i)\, \sin\, \left[ \frac{2 \pi}{T}*k
                            + {2 \phi(i)} \right] + b(i)
        \label{farad}\,,
   \end{equation}

   where $k$ is the channel number, $i$ is the pulse longitude,
   $T$ is a period of a sinusoidal curve measured in channels,
   $\phi(i)$ is a position angle profile, and $a(i)$ is an amplitude
   of sinusoidal modulation of the polarized component.
   The computed values of $A_{k}(i)$ give the Faraday rotation
   spectrum of the average profile and corresponding pulse shapes
   in each channel. The results are presented in Fig.~\ref{model}.

   In computing $A_{k}(i)$ we used the published polarization
   profiles of PSR B0950+08 at low frequencies at 150 and 151 MHz
   given in Schwarz \& Morris (\cite{schwarz}) and
   Lyne et al. (\cite{lyne}). The model profile adopted is presented
   in Fig.~\ref{model}a that shows the average profile in total
   intensity $I(i)$, the linearly polarized component $P(i)$, and
   the position angle of polarization $\phi(i)$. The profile for $I(i)$
   is normalized to 100 at maximum. In this model, the unresolved
   component ahead of the main pulse is 100$\%$ linearly polarized
   (longitude interval from 1$\degr$ to 29$\degr$).
   The degree of linear polarization, $P_{\mathrm{L}}$, for the first
   and second components of the main pulse is 75$\%$ and 80$\%$,
   respectively. Their corresponding longitude intervals are
   29$\degr$--44$\degr$ and 44$\degr$--66$\degr$.

   The amplitude of sinusoidal curve $a(i)$ in Eq.~(\ref{farad})
   was defined by the relation $a(i)=P(i)/2$,
   where $P(i)=I(i)*P_{\mathrm{L}}(i)$ is the intensity of
   the polarized component at longitude $i$.
   The mean level of the pulse intensity $b(i)$ at
   longitude $i$ was found from $b(i)=I(i)/2$. The period of
   the sinusoidal curve was assumed equal to  $T$=256 channels,
   the same as in  Fig.~\ref{sin256}. According to
   Lyne et al. (\cite{lyne}), the position angle $\phi$ swings
   160$\degr$ through the profile. This gives the phase offset
   between the Faraday spectra of two components of the main pulse
   of about -0.4 radian, i.e. the same as in Eq.~(\ref{y2}).
   To obtain the phase offset between the spectra of
   the first and third components equal to
   $\approx {\mathrm{\pi}}$, as specified in Eq.~(\ref{y3})
   and Fig.~\ref{sin256}, we should suppose that
   the position angles for these components are orthogonal.
   The finally adopted behavior of the position angle is given in
   Fig.~\ref{model}a in the top panel. Here the position
   angle swings 200$\degr$ through the profile and decreases
   across the profile with a higher rate of decrease in the range
   of the third component. This gives about 90$\degr$ displacement
   from the position angle of the first component.

   The resulting frequency dependence of the pulse profile at
   the longitudes of the pulse peaks of the three components
   is given in Fig.~\ref{model}b.
   Comparison with Fig.~\ref{sin256} shows that the computed
   Faraday rotation spectra are generally in good agreement with
   the observed frequency dependence of the profile.
   The values of the phase offsets between the three sinusoidal
   curves are identical in both plots.
   The values of the amplitudes are slightly different.
   The amplitudes of the observed sinusoidal curves for the two
   components of the main pulse are about half that shown
   in Fig.~\ref{model}b. Possibly, the procedure
   of removing the effects of scintillations,
   which was mentioned in \S5, could partially decrease
   the amplitude of the wave, unrelated to scintillations.
   The amplitude of the third component ahead of the main pulse
   is 4 times greater than that plotted in Fig.~\ref{model}b.
   Possibly, at 112 MHz this component is stronger than
   it was assumed in our model. We do not know the true
   profile in total intensity at 112 MHz and therefore the assumed
   ratios of amplitudes in the various profile components
   are approximate.

   The corresponding profile shapes computed for individual
   channels are given in Fig.~\ref{model}c. Comparison with
   Fig.~\ref{prof5khz} shows that the computed
   profiles are in good agreement with the observed ones.
   We clearly see the noticeable feature which characterizes
   the observed modulation of the pulse profiles.
   This is a transition state between the first and second dominant
   component of the main pulse. This state has a lower S/N ratio
   and shows the presence of the well-resolved third component
   ahead of the main pulse. The well-resolved third
   component does not exist. The transformation of the unresolved
   component of the pulse profile to the well-resolved one
   is a result of polarization effects.
   The simulation of the transition state was successful
   when the degree of linear polarization taken
   was very high, more than 80$\%$.
   Hence, we conclude that the observed profile changes
   are a result of the Faraday rotation effect.

\section{Narrowband changes of the average profile
         at lower frequencies of 88, 62 and 41 MHz}

   Low-frequency observations of the pulsar were carried out
   to examine sinusoidal modulation of the pulse profile in
   the wide frequency range and to estimate the polarization
   effects corresponding to our observations. As mentioned before,
   the observed effects at 112 MHz are about 3 times stronger
   than those predicted by the interstellar Faraday rotation.

\begin{figure}
   \centering
   \includegraphics[width=6cm]{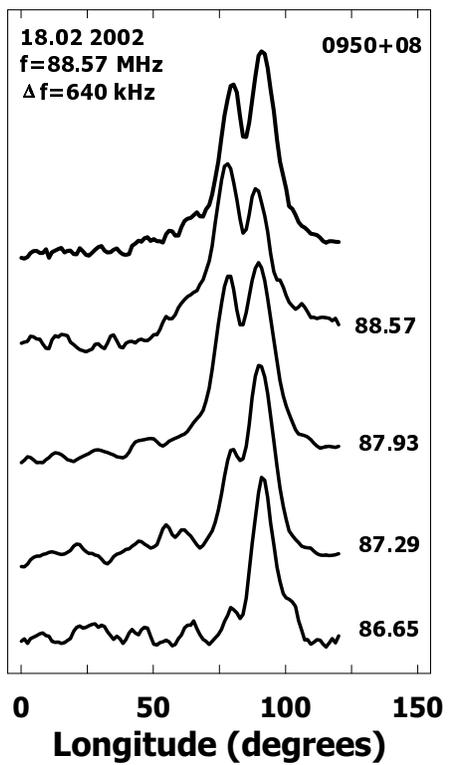}
   \caption{
    The pulse profile changes at the frequency of 88.57 MHz.
    The upper profile is a sum of 3700 pulsar periods over a total
    2.56-MHz band. The lower profiles are that over four 640-kHz bands.
    The sample interval is $1{\fdg}2$ of longitude.
    All the profiles are normalized to unity.
         }
   \label{prf88mhz}
   \end{figure}

   Fig.~\ref{prf88mhz} displays the average pulse profile
   at 88.57 MHz. The profile exhibits
   a well-resolved double shape with a separation between
   the components of about 11$\degr$. As we can see from this plot,
   the profile shape varies markedly across the bandpass and
   these changes are similar to the changes observed at
   the higher frequency of 112 MHz.
   The transition state between the first and second
   dominant components was not recorded during this
   observation, but one can guess its presence.
   The frequency interval of the shape changes can be estimated
   to be about 3 MHz.

\begin{figure}
   \centering
   \includegraphics[width=6cm]{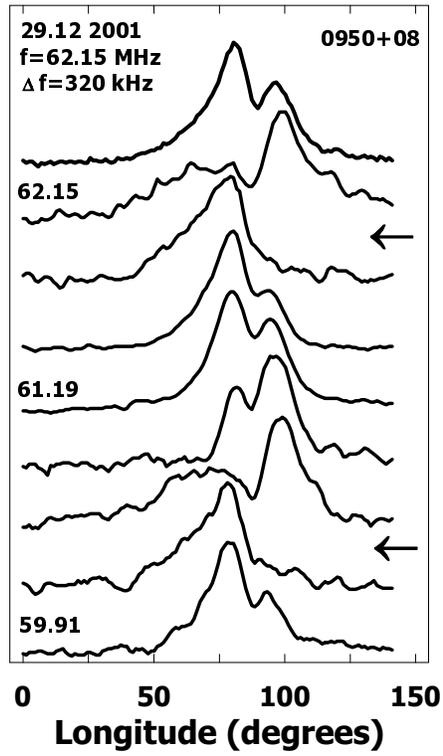}
   \caption{
    The pulse profile changes at the frequency of 62.15 MHz.
    The upper profile is a sum of 2200 pulsar periods over a total
    2.56-MHz band. The lower profiles are that over
    eight 320-kHz bands. A transition state between the first and
    second dominant components is marked by an arrow.
    The sample interval is $1{\fdg}2$ of longitude.
    All the profiles are normalized to unity.
         }
   \label{prf62mhz}
   \end{figure}

   A similar picture of the narrowband changes of the pulse profile
   is observed at 62.15 MHz as is shown in Fig.~\ref{prf62mhz}.
   Here the profile is obtained by addition of 2200 pulse periods
   within eight 320-kHz bands. The average profile has
   two well-resolved components with a separation of 15$\degr$.
   Obviously, the profile shape varies across the bandpass
   and the frequency structure of the profile is similar to that
   observed at higher frequencies of 88 and 112 MHz.
   A noticeable feature of the shape changes is the presence of
   two transition states between the first and second dominant
   components, which are clearly visible in the plot.
   The frequency interval of the profile changes of
   approximately 1.2 MHz.

\begin{figure}
   \centering
   \includegraphics[width=8cm]{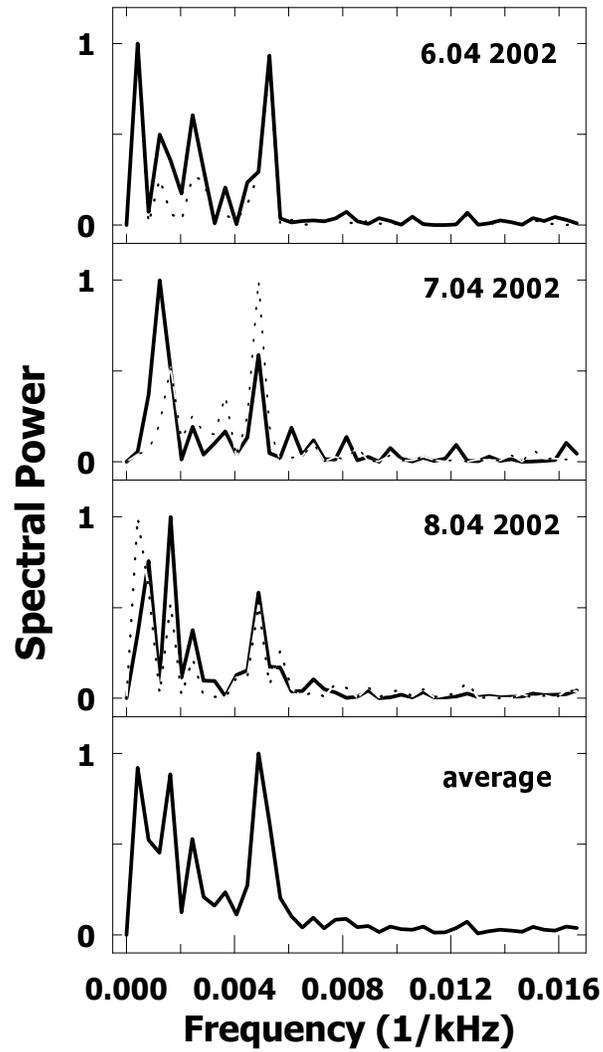}
   \caption{
    The power spectra of the amplitude variations of individual
    pulses for PSR B0950+08 at 41 MHz for three different observations.
    The spectral feature at around 0.005 ${\mathrm{kHz^{-1}}}$
    is well visible in the spectra for all the three observations.
    In the average spectrum shown in the lower panel this feature
    dominates over the noisy fluctuations.
    }
   \label{prf41mhz}
   \end{figure}

   Average profiles of the pulsar at 41.07 MHz do not show
   a frequency structure similar to that observed at higher
   frequencies. The profiles have a lower S/N ratio and much lower
   time resolution of 3 ms, caused by the 7-ms pulse
   broadening due to the dispersion effect in one 20-kHz band.
   The profiles would be expected to be modulated by
   the interstellar Faraday rotation with the frequency scale
   of the order of 900 kHz. To find periodicities
   in the amplitude variation of the pulse profile over
   the bandpass of 2.56 MHz, the power spectrum was computed
   with the use of Fourier transform.
   For this procedure, we used strong individual pulses.
   It is well known that individual pulses are often more
   highly polarized than average profiles.
   The results are given in Fig.~\ref{prf41mhz} for three
   different epochs. For each observation two realizations
   of the power spectrum are presented.

   There is a distinct spectral feature at around
   0.005 ${\mathrm{kHz^{-1}}}$, which is clearly visible in
   the spectra for the three observations. This feature
   corresponds to the periodicity of approximately 200 kHz.
   To state with certainty the presence of this spectral feature
   we averaged the spectra together and presented the resulting
   spectrum in the lower panel of Fig.~\ref{prf41mhz}. As is seen,
   the spectral feature at 0.005 ${\mathrm{kHz^{-1}}}$
   dominates over the noisy fluctuations.
   This is a strong indication of the presence of sinusoidal
   modulation of the pulse profile at frequency of 41 MHz, too.

   In this section we analyzed the narrowband profile
   changes in the low-frequency range from 41 to 88 MHz and concluded
   that the shape changes are similar to those observed at 112 MHz.
   The scale of the observed profile changes is frequency dependent
   and varies from 3--9 MHz at 112 MHz to 0.2 MHz at 41 MHz.

\section{Timing residuals for PSR B0950+08}

   The pulsar catalog (Taylor et al. \cite{taylor}) quotes the period
   and the period derivative obtained by Gullahorn \& Rankin (1978)
   at the epoch of 1972.5. Errors in these parameters alone
   lead to ambiguities in the counts of the pulse number for
   the 1996--2001 observing session.
   Our measurements of the average profiles were used
   to improve the timing parameters for PSR B0950+08.

   We analyzed the arrival times for only those average profiles
   whose shape looks double. They were chosen out of the full
   data set obtained at 102.5 and 112 MHz between 1996 and 2001.
   All good-quality double profiles were summed to produce
   an average profile with a high S/N ratio which was used
   as a template. The topocentric arrival times were calculated
   by cross-correlating the observed profiles with this template.
   Timing parameters were determined using the TEMPO software
   package\footnote
   {http://www.atnf.csiro.au/research/pulsar/timing/tempo}
   and the JPL DE200 ephemeris.
   The topocentric arrival times were referred to the solar
   system barycenter using a position and a proper motion given
   by Fomalont et al. (\cite{fomalont}) and Brisken et al.
   (\cite{brisken}), respectively.
   A second-order polynomial describing spin-down pulsar behavior
   was fitted to the barycentric arrival times to obtain residuals
   from a timing model. Residuals were used to compute
   differential corrections to the initial parameter values.
%
   \begin{table}
      \caption[]{Measured timing parameters for PSR B0950+08}
         \label{param}
    $$
         \begin{tabular}{ll}
            \hline
            \noalign{\smallskip}
Parameter                            &    Value$^{\mathrm{a}}$   \\
            \noalign{\smallskip}
            \hline
            \noalign{\smallskip}
Period, $P\,({\mathrm{s}})$               & 0.253065240711(4)    \\
Period derivative,
  $\dot{P}\ (10^{-15}\,{\mathrm{s}}\, {\mathrm{s}^{-1}})$
                                          & 0.23028(5)           \\
Epoch of period (MJD)                     & 50190.7360           \\
            \noalign{\smallskip}
            \hline
         \end{tabular}
    $$
\begin{list}{}{}
\item[$^{\mathrm{a}}$] The errors are given in units of the
                       last quoted digit
\end{list}
   \end{table}

\begin{figure}
   \includegraphics[width=11cm]{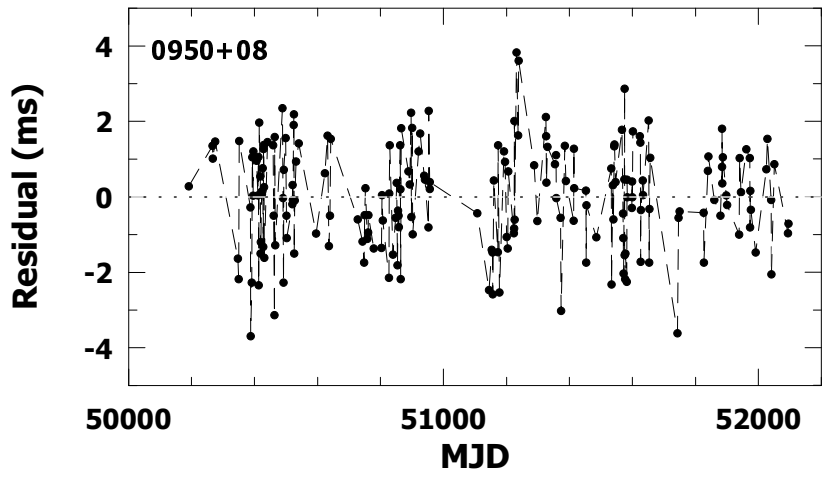}
   \caption{
    Post-fit timing residuals for PSR B0950+08 between 1996 April
    and 2001 July.
           }
   \label{resid}
   \end{figure}

   The best-fitting values of the period and the period
   derivative are presented in Table~\ref{param}.
   A large scatter of timing data relative to the spin-down pulsar
   model has not allowed us to improve the astrometric parameters.
   Fig.~\ref{resid} displays the timing residuals from
   the pulsar relative to their best-fit spin-down model
   during the period 1996--2001.
   The rms timing residual after the fit was of the order of
   1300 ${\mu}s$ with a timing error of about 90 ${\mu}s$.
   At higher frequencies, above 400 MHz, where the pulsar
   exhibits a single pulse profile, the rms timing residual
   of PSR B0950+08 is much less and is estimated at
   200 ${\mu}s$ with a timing uncertainty of the order of
   90--200 ${\mu}s$ in a 4--7 year span of data
   (Gullahorn \& Rankin \cite{gullahorn};
   Helfand et al. \cite{helfand1}).
   Apparently, the increase of the rms residual observed at
   112 MHz is due to the profile shape changes which cause
   fluctuations of the pulse arrival times and make an additional
   noise contribution to the timing residual.

\section{Discussion}

   We have found that the average pulse profile for PSR B0950+08
   observed with a single linear polarization
   is frequency variable all over the range 41--112 MHz.
   The narrowband structure of the pulse shape is associated
   with sinusoidal modulation of the pulsar emission.
   This process is frequency dependent.
   According to our measurements, the interval of modulation
   at different observing frequencies is about
   3--9 MHz at 111.87 MHz, 3--4 MHz at 88.57 MHz, 1--1.4 MHz
   at 62.15 MHz and 0.19--0.21 MHz at 41.07 MHz.
   Hence, the dependence of the modulation interval on frequency
   is well described by a power low with index close to 3 in
   the range  41--112 MHz. A similar frequency structure should
   be observed in the presence of the Faraday rotation effect.
   The presence of Faraday rotation is also confirmed by
   a numerical simulation of the polarization effects
   for the pulse profiles at 112 MHz.

   The frequency interval of Faraday rotation of the plane
   of polarization of linear polarized emission
   ${\Delta}F_{\mathrm{\pi}}$ at the observing frequency
   ${\nu}$ is determined by rotation in the position angle
   ${\theta}$ through  ${\pi}$ and is calculated from

   \begin{equation}
       {\Delta}F_{\mathrm{\pi}} =
                  \frac{{\pi}\, {\nu}^{3}}{2\, RM\, c^{2}}
                       = 17.48\, {\nu}^{3}\, / \, RM
       \label{fpi} \, ,
   \end{equation}

   where ${\Delta}F_{\mathrm{\pi}}$ is in MHz, ${\nu}$ is in
   hundreds of MHz, the rotation measure $RM$ is in
   ${\mathrm{rad\, m^{-2}}}$ and $c$ is the light velocity.

   Using relation~(\ref{fpi}), we can estimate the rotation
   measure $RM$ corresponding to our measurements.
   For the mean values of ${\Delta}F_{\mathrm{\pi}}$
   equal to 6, 3, 1.2 and 0.21 MHz at the frequencies of
   111.87, 88.57, 62.15 and 41.07 MHz respectively, the
   rotation measure $RM$ will be approximately
   4 ${\mathrm{rad\, m^{-2}}}$. Taking into account
   variability of ${\Delta}F_{\mathrm{\pi}}$ from one
   observation to another, we derive the values of $RM$ in
   the range 3--6 ${\mathrm{rad\, m^{-2}}}$.
   The derived values of $RM$ are very large compared to
   the rotation measure $RM = 1.35\ {\mathrm{rad\, m^{-2}}}$
   measured for PSR B0950+08 at frequencies above 400 MHz
   (Hamilton \& Lyne \cite{hamilton}; Taylor et al. \cite{taylor}).

   Thus, we have shown that the observed narrowband profile changes
   of the pulsar B0950+08 are a result of the Faraday rotation effect
   and could be explained as an artifact of observations of linearly
   polarized pulsar emission with a linearly polarized antenna.
   We have obtained a new value of the rotation measure corresponding
   to this effect, $RM \approx 4\ {\mathrm{rad\, m^{-2}}}$.
   This leads us to conclude that the published value of $RM$
   is incorrect and is actually 3 times greater.

\begin{acknowledgements}
      We wish to thank the staff of PRAO for assistance in
      the observations. This work was supported by
      the Russian Foundation For Basic Research
      (project No. 00-02-17447).
      The authors are grateful to the referee for numerous
      helpful comments and suggestions.

\end{acknowledgements}

%
   \newpage
   \clearpage
%
%
   \newpage
   \clearpage
%
   \newpage
   \clearpage
%
%
   \newpage
   \clearpage
%
%
   \newpage
   \clearpage
%
   \newpage
   \clearpage
%
   \newpage
   \clearpage
%
   \newpage
   \clearpage
%
   \newpage
   \clearpage
%
   \newpage
   \clearpage
\end{document}